%Paper: gr-qc/9507015
%From: ESPOSITO@napoli.infn.it
%Date: Thu, 6 Jul 1995 17:48:39 +0200 (CET-DST)

\magnification \magstep1
\raggedbottom
\openup 4\jot
\voffset6truemm
\headline={\ifnum\pageno=1\hfill\else
\hfill {\it Twistors in conformally flat Einstein
four-manifolds} \hfill \fi}
\def\cstok#1{\leavevmode\thinspace\hbox{\vrule\vtop{\vbox{\hrule\kern1pt
\hbox{\vphantom{\tt/}\thinspace{\tt#1}\thinspace}}
\kern1pt\hrule}\vrule}\thinspace}
\centerline {\bf TWISTORS IN CONFORMALLY FLAT}
\centerline {\bf EINSTEIN FOUR-MANIFOLDS}
\vskip 0.3cm
\centerline {Giampiero Esposito$^{1,2}$
and Giuseppe Pollifrone$^{3}$}
\vskip 0.3cm
\centerline {\it ${ }^{1}$Istituto Nazionale di Fisica Nucleare,
Sezione di Napoli}
\centerline {\it Mostra d'Oltremare Padiglione 20,
80125 Napoli, Italy;}
\centerline {\it ${ }^{2}$Dipartimento di Scienze Fisiche}
\centerline {\it Mostra d'Oltremare Padiglione 19,
80125 Napoli, Italy;}
\centerline {\it ${ }^{3}$Dipartimento di Fisica, Universit\`a
di Roma ``La Sapienza"}
\centerline {\it and INFN, Sezione di Roma,
Piazzale Aldo Moro 2, 00185 Roma, Italy.}
\vskip 0.3cm
\noindent
{\bf Abstract}. This paper studies the two-component spinor
form of massive spin-${3\over 2}$ potentials in conformally
flat Einstein four-manifolds. Following earlier work in the
literature, a non-vanishing cosmological constant makes it
necessary to introduce a supercovariant derivative operator.
The analysis of supergauge transformations
of primary and secondary potentials
for spin ${3\over 2}$ shows that the gauge freedom for massive
spin-${3\over 2}$ potentials is generated by solutions of the
supertwistor equations. The supercovariant form of a partial
connection on a non-linear bundle is then obtained, and
the basic equation of massive secondary potentials is shown
to be the integrability condition on super $\beta$-surfaces of a
differential operator on a vector bundle of rank three.
Moreover, in the presence of boundaries,
a simple algebraic
relation among some spinor fields is found
to ensure the gauge invariance of locally
supersymmetric boundary conditions
relevant for quantum cosmology and supergravity.
\vskip 0.3cm
\leftline {PACS numbers: 0420, 0465, 9880}
\vskip 1cm
\leftline {\bf 1. Introduction}
\vskip 1cm
\noindent
The local theory of spin-${3\over 2}$ potentials in real Riemannian
4-geometries is receiving careful consideration in the current
literature. There are at least two motivations for this analysis. In
Minkowski space-time, twistors arise naturally as charges for massless
spin-${3\over 2}$ fields [1-5]. In Ricci-flat 4-manifolds, such fields are
well defined (Ricci-flatness being a necessary and sufficient consistency
condition), and a suitable generalization of the concept of twistors
would make it possible to reconstruct solutions of the vacuum
Einstein equations out of the resulting twistor space. In
extended supergravity theories, however, it is necessary to make
sense of massive spin-${3\over 2}$ fields in Riemannian backgrounds.
For this purpose, a careful spinorial analysis of the
problem is in order.

We have thus focused on massive spin-${3\over 2}$ potentials in
4-manifolds with non-vanishing cosmological constant, considering the
supercovariant derivative compatible with a non-vanishing scalar
curvature. This is the content of section 2. Section 3 studies the
gauge freedom of the second kind, which is generated by a particular
type of twistors, i.e. the Euclidean Killing spinors. Section 4
studies the preservation of spin-${3\over 2}$ field equations under
the supergauge transformations of primary potentials. Section 5
studies secondary potentials for spin ${3\over 2}$ in the massive
case. In section 6 a partial
superconnection acting on a bundle
with non-linear fibres is introduced.
Section 7 studies
the action of a superconnection
on a vector bundle of rank three,
and the corresponding integrability condition on
super $\beta$-surfaces is derived.
Section 8 studies the case of backgrounds
with boundaries and shows the conditions under which
locally supersymmetric boundary conditions
are gauge-invariant.
Results and open problems are described in section 9. Relevant
details are given in the appendix.
\vskip 1cm
\leftline {\bf 2. The superconnection}
\vskip 1cm
\noindent
In the massless case, the two-spinor form of the
Rarita-Schwinger equations is the one given in the appendix,
where $\nabla_{AA'}$ is the spinor covariant derivative
corresponding to the connection $\nabla$ of the background.
In the massive case, however, the appropriate connection,
hereafter denoted by $S$, has an additional term which couples
to the cosmological constant $\lambda=6\Lambda$
[9,10]. In the language of
$\gamma$-matrices, the new covariant derivative $S_{\mu}$ to be
inserted {\it in the field equations} takes the form [9,10]
$$
S_{\mu} \equiv \nabla_{\mu}+f(\Lambda)\gamma_{\mu}
\eqno (2.1)
$$
where $f(\Lambda)$ vanishes at $\Lambda=0$, and $\gamma_{\mu}$
are the curved-space $\gamma$-matrices. Since, following [1-8],
we are interested in the two-spinor formulation of the problem,
we have to bear in mind the action of $\gamma$-matrices on any
spinor $\varphi \equiv \Bigr(\beta^{C},
{\widetilde \beta}_{C'}\Bigr)$. Note that unprimed and primed
spin-spaces are no longer (anti)-isomorphic in the case of
positive-definite 4-metrics, since there is no complex
conjugation which turns primed spinors into unprimed spinors
(or the other way around) [5,11]. Hence $\beta^{C}$ and
${\widetilde \beta}_{C'}$ are totally unrelated. With this
understanding, we write the supergauge transformations for
massive spin-${3\over 2}$ potentials in the form
(cf [1-5])
$$
{\widehat \gamma}_{\; \; B'C'}^{A} \equiv
\gamma_{\; \; B'C'}^{A}+S_{\; \; B'}^{A} \; \lambda_{C'}
\eqno (2.2)
$$
$$
{\widehat \Gamma}_{\; \; \; BC}^{A'} \equiv
\Gamma_{\; \; \; BC}^{A'}+S_{\; \; \; B}^{A'} \;
\nu_{C}
\eqno (2.3)
$$
where the action of $S_{AA'}$ on the gauge fields
$\Bigr(\nu^{B},\lambda_{B'}\Bigr)$ is defined by (cf (2.1))
$$
S_{AA'} \; \nu_{B} \equiv \nabla_{AA'} \; \nu_{B}
+f_{1}(\Lambda)\epsilon_{AB} \; \lambda_{A'}
\eqno (2.4)
$$
$$
S_{AA'} \; \lambda_{B'} \equiv \nabla_{AA'} \; \lambda_{B'}
+f_{2}(\Lambda) \epsilon_{A'B'} \; \nu_{A} .
\eqno (2.5)
$$
With our notation, $R=24 \Lambda$ is the scalar curvature,
$f_{1}$ and $f_{2}$ are two functions which vanish at
$\Lambda=0$, whose form will be determined later by a
geometric analysis. The action of $S_{AA'}$ on a many-index
spinor $T_{B'...F'}^{A...L}$ can be obtained by expanding
such a $T$ as a sum of products of spin-vectors, i.e. [12]
$$
T_{B'...F'}^{A...L}=\sum_{i} \alpha_{(i)}^{A} ...
\beta_{(i)}^{L} \; \gamma_{B'}^{(i)} ... \delta_{F'}^{(i)}
\eqno (2.6)
$$
and then applying the Leibniz rule and the definitions
(2.4)-(2.5), where $\alpha_{(i)}^{A}$ has an independent
partner ${\widetilde \alpha}_{(i)}^{A'}$, ... ,
$\gamma_{B'}^{(i)}$ has an independent partner
${\widetilde \gamma}_{B}^{(i)}$, ... , and so on. A further,
non-trivial requirement is that $S_{AA'}$ should annihilate
the curved $\epsilon$-spinors [12], in much the same way as
$\nabla_{AA'}$ annihilates such spinors. In our analysis we
always assume that
$$
S_{AA'} \; \epsilon_{BC}=0
\eqno (2.7)
$$
$$
S_{AA'} \; \epsilon_{B'C'}=0 .
\eqno (2.8)
$$
In the light of the definitions and assumptions presented
so far, one can make sense of the Rarita-Schwinger equations
with non-vanishing cosmological constant
$\lambda=6\Lambda$, i.e. (cf appendix)
$$
\epsilon^{B'C'} \; S_{A(A'} \;
\gamma_{\; \; B')C'}^{A}=\Lambda \; {\widetilde F}_{A'}
\eqno (2.9)
$$
$$
S^{B'(B} \; \gamma_{\; \; \; B'C'}^{A)}=0
\eqno (2.10)
$$
$$
\epsilon^{BC} \; S_{A'(A} \; \Gamma_{\; \; \; B)C}^{A'}
=\Lambda \; F_{A}
\eqno (2.11)
$$
$$
S^{B(B'} \; \Gamma_{\; \; \; \; BC}^{A')}=0 .
\eqno (2.12)
$$
With our notation, $F_{A}$ and ${\widetilde F}_{A'}$ are
spinor fields proportional to the traces of secondary
potentials for spin ${3\over 2}$. These will be studied in
section 5.
\vskip 1cm
\leftline {\bf 3. Gauge freedom of the second kind}
\vskip 1cm
\noindent
The gauge freedom of the second kind is the one which does
not affect the potentials after a gauge
transformation. This
requirement corresponds to the case analyzed in [13],
where it is pointed out that whilst the Lagrangian of
$N=1$ supergravity is invariant under gauge transformations
with arbitrary spinor fields $\Bigr(\nu^{A},\lambda_{A'}\Bigr)$,
the actual {\it solutions} are only invariant if the
supercovariant derivatives (2.4)-(2.5) vanish.

On setting to zero $S_{AA'} \; \nu_{B}$ and
$S_{AA'} \; \lambda_{B'}$, one gets a coupled set of
equations which are the Euclidean version of the
Killing-spinor equation [13], i.e.
$$
\nabla_{\; \; \; B}^{A'} \; \nu_{C}=-f_{1}(\Lambda)
\lambda^{A'} \; \epsilon_{BC}
\eqno (3.1)
$$
$$
\nabla_{\; \; B'}^{A} \; \lambda_{C'}=-f_{2}(\Lambda)
\nu^{A} \; \epsilon_{B'C'} .
\eqno (3.2)
$$
What is peculiar of equations (3.1)-(3.2) is that their
right-hand sides involve spinor fields which are,
themselves, solutions of the twistor equation. Hence one
deals with a special type of twistors, which do not exist
in a generic curved manifold (cf [13]). Equation (3.1) can
be solved for $\lambda^{A'}$ as
$$
\lambda_{C'}={1\over 2f_{1}(\Lambda)}\nabla_{C'}^{\; \; \; B}
\; \nu_{B} .
\eqno (3.3)
$$
The insertion of (3.3) into (3.2) and the use of spinor
Ricci identities [5,12] yields the second-order equation
$$
\cstok{\ }\nu_{A}+(6\Lambda+8f_{1}f_{2})\nu_{A}=0 .
\eqno (3.4)
$$
On the other hand, (3.1) implies the twistor equation
$$
\nabla_{\; \; \; (B}^{A'} \; \nu_{C)}=0 .
\eqno (3.5)
$$
Covariant differentiation of (3.5), jointly with spinor Ricci
identities, leads to [8]
$$
\cstok{\ }\nu_{A}-2\Lambda \nu_{A}=0 .
\eqno (3.6)
$$
By comparison of (3.4) and (3.6) one finds the condition
$f_{1}f_{2}=-\Lambda$. The integrability condition of (3.5)
is given by [11]
$$
\psi_{ABCD} \; \nu^{D}=0 .
\eqno (3.7)
$$
This means that our manifold is conformally left-flat, unless
$\nu^{D}$ is a four-fold principal spinor of the
anti-self-dual Weyl spinor. The latter possibility is here
ruled out, to avoid having gauge fields related explicitly
to the curvature of the background.

The condition $f_{1}f_{2}=-\Lambda$ is also obtained by
comparison of first-order equations, since for example
$$
\nabla^{AA'} \; \nu_{A}=2f_{1}\lambda^{A'}
=-2{\Lambda \over f_{2}} \lambda^{A'} .
\eqno (3.8)
$$
The first equality in (3.8) results from (3.1), whilst the
second one is obtained since the twistor equations also
imply that (see (3.2))
$$
\nabla^{AA'} \Bigr(-f_{2}\nu_{A}\Bigr)
=2\Lambda \; \lambda^{A'} .
\eqno (3.9)
$$
Entirely analogous results are obtained on considering the
twistor equation resulting from (3.2), i.e.
$$
\nabla_{\; \; (B'}^{A} \; \lambda_{C')}=0 .
\eqno (3.10)
$$
The integrability condition of (3.10) is
$$
{\widetilde \psi}_{A'B'C'D'} \; \lambda^{D'}=0 .
\eqno (3.11)
$$
Since our gauge fields are not assumed to be four-fold principal
spinors of the Weyl spinor (cf [14]), equations (3.7) and (3.11)
imply that our background geometry is conformally flat.
\vskip 1cm
\leftline {\bf 4. Compatibility conditions}
\vskip 1cm
\noindent
We now require that the field equations (2.9)-(2.12) should
be preserved under the action of the supergauge transformations
(2.2)-(2.3). This is the procedure one follows in the massless
case, and is a milder requirement with respect to the analysis
of section 3.

If $\nu^{B}$ and $\lambda_{B'}$ are twistors, but not necessarily
Killing spinors, they obey the equations (3.5) and (3.10), which
imply that, for some independent spinor fields $\pi^{A}$ and
${\widetilde \pi}^{A'}$, one has
$$
\nabla_{\; \; \; B}^{A'} \; \nu_{C}
=\epsilon_{BC} \; {\widetilde \pi}^{A'}
\eqno (4.1)
$$
$$
\nabla_{\; \; B'}^{A} \; \lambda_{C'}
=\epsilon_{B'C'} \; \pi^{A} .
\eqno (4.2)
$$
In the compatibility equations, whenever one has terms of the
kind $S_{AA'} \; \nabla_{\; \; B'}^{A} \; \lambda_{C'}$, it is
therefore more convenient to symmetrize and anti-symmetrize over
$B'$ and $C'$. A repeated use of this algorithm leads to a
considerable simplification of the lengthy calculations. For
example, the preservation condition of (2.9) has the
general form
$$
3f_{2}\Bigr(\nabla_{AA'} \; \nu^{A}+2f_{1}\lambda_{A'}\Bigr)
+\epsilon^{B'C'}\biggr[S_{AA'}\Bigr(\nabla_{\; \; B'}^{A}
\; \lambda_{C'}\Bigr)+S_{AB'}\Bigr(\nabla_{\; \; A'}^{A}
\; \lambda_{C'}\Bigr)\biggr]=0 .
\eqno (4.3)
$$
By virtue of (4.2), equation (4.3) becomes
$$
f_{2}\Bigr(\nabla_{AA'} \; \nu^{A}+2f_{1}\lambda_{A'}\Bigr)
+S_{AA'} \; \pi^{A}=0 .
\eqno (4.4)
$$
Following (2.4)-(2.5), the action of the
supercovariant derivative on $\pi_{A},{\widetilde \pi}_{A'}$
yields
$$
S_{AA'} \; \pi_{B} \equiv \nabla_{AA'} \; \pi_{B}
+f_{1}(\Lambda)\epsilon_{AB} \; {\widetilde \pi}_{A'}
\eqno (4.5)
$$
$$
S_{AA'} \; {\widetilde \pi}_{B'} \equiv
\nabla_{AA'} \; {\widetilde \pi}_{B'}
+f_{2}(\Lambda)\epsilon_{A'B'} \; \pi_{A} .
\eqno (4.6)
$$
Equations (4.4)-(4.5), jointly with the equations
$$
\cstok{\ }\lambda_{A'}-2\Lambda \; \lambda_{A'}=0
\eqno (4.7)
$$
$$
\nabla^{AA'} \; \pi_{A}=2\Lambda \; \lambda^{A'}
\eqno (4.8)
$$
which result from (4.2), lead to
$$
(f_{1}+f_{2}){\widetilde \pi}_{A'}
+(f_{1}f_{2}-\Lambda)\lambda_{A'}=0 .
\eqno (4.9)
$$
Moreover, the preservation of (2.10) under (2.2) leads to
the equation
$$
S^{B'(A} \; \pi^{B)} + f_{2} \nabla^{B'(A} \;
\nu^{B)}=0
\eqno (4.10)
$$
which reduces to
$$
\nabla^{B'(A} \; \pi^{B)}=0
\eqno (4.11)
$$
by virtue of (4.1) and (4.5). Note that a supertwistor is
also a twistor, since
$$
S^{B'(A} \; \pi^{B)}=\nabla^{B'(A} \; \pi^{B)}
\eqno (4.12)
$$
by virtue of the definition (4.5). It is now clear that,
for a gauge freedom generated by twistors, the preservation
of (2.11)-(2.12) under (2.3) leads to the compatibility
equations
$$
(f_{1}+f_{2})\pi_{A}+(f_{1}f_{2}-\Lambda)\nu_{A}=0
\eqno (4.13)
$$
$$
\nabla^{B(A'} \; {\widetilde \pi}^{B')}=0
\eqno (4.14)
$$
where we have also used the equation (see (3.6) and (4.1))
$$
\nabla^{AA'} \; {\widetilde \pi}_{A'}=2\Lambda \; \nu^{A} .
\eqno (4.15)
$$
Note that, if $f_{1}+f_{2} \not = 0$,
one can solve (4.9) and (4.13) as
$$
{\widetilde \pi}_{A'}={(\Lambda-f_{1}f_{2})\over (f_{1}+f_{2})}
\lambda_{A'}
\eqno (4.16)
$$
$$
\pi_{A}={(\Lambda-f_{1}f_{2})\over (f_{1}+f_{2})}\nu_{A}
\eqno (4.17)
$$
and hence one deals again with Euclidean Killing spinors as
in section 3. However, if
$$
f_{1}+f_{2}=0
\eqno (4.18)
$$
$$
f_{1}f_{2}-\Lambda=0
\eqno (4.19)
$$
the spinor fields ${\widetilde \pi}_{A'}$ and $\lambda_{A'}$
become {\it unrelated}, as well as $\pi_{A}$ and $\nu_{A}$.
This is a crucial point. Hence one may have
$f_{1}=\pm \sqrt{-\Lambda}$, $f_{2}=\mp \sqrt{-\Lambda}$, and
one finds a more general structure.

In the generic case, we do not assume that $\nu^{B}$ and
$\lambda_{B'}$ obey any equation. This means that, on the
second line of equation (4.3), it is more convenient to
express the term in square brackets as
$2S_{A(A'} \; \nabla_{\; \; B')}^{A} \; \lambda_{C'}$. The
rules of section 2 for the action of $S_{AA'}$ on spinors with
many indices lead therefore to the compatibility
conditions
$$
3f_{2} \Bigr(\nabla_{AA'} \; \nu^{A}+2f_{1}\lambda_{A'}
\Bigr)-6\Lambda \; \lambda_{A'}
+4f_{1}{\widetilde P}_{(A'B')}^{\; \; \; \; \; \; \; \; B'}
+3f_{2}{\widetilde Q}_{A'}=0
\eqno (4.20)
$$
$$
3f_{1} \Bigr(\nabla_{AA'} \; \lambda^{A'}
+2f_{2} \nu_{A}\Bigr)-6\Lambda \; \nu_{A}
+4f_{2}P_{(AB)}^{\; \; \; \; \; \; B}
+3f_{1}Q_{A}=0
\eqno (4.21)
$$
$$
\Phi_{\; \; \; \; C'D'}^{AB} \; \lambda^{D'}
+f_{2}U_{\; \; \; \; \; \; C'}^{(AB)}
-f_{2}\nabla_{C'}^{\; \; \; (A} \; \nu^{B)}=0
\eqno (4.22)
$$
$$
{\widetilde \Phi}_{\; \; \; \; \; \; CD}^{A'B'} \; \nu^{D}
+f_{1}{\widetilde U}_{\; \; \; \; \; \; \; \; C}^{(A'B')}
-f_{1}\nabla_{C}^{\; \; (A'} \; \lambda^{B')}=0
\eqno (4.23)
$$
where the detailed form of
$P,{\widetilde P},Q,{\widetilde Q}$ is not strictly necessary,
but we can say that they do not depend explicitly on the
trace-free part of the Ricci spinor, or on the Weyl spinors.
Note that, in the massless limit $f_{1}=f_{2}=0$, the equations
(4.20)-(4.23) reduce to the familiar form of compatibility
equations which admit non-trivial solutions only in Ricci-flat
backgrounds [8].

Our consistency analysis still makes it necessary to set to
zero $\Phi_{\; \; \; \; C'D'}^{AB}$ (and hence
${\widetilde \Phi}_{\; \; \; \; \; \; CD}^{A'B'}$ by reality
[11]). The remaining contributions to (4.20)-(4.23) should then
become algebraic relations by virtue of the twistor equation.
This is confirmed by the analysis of gauge freedom for
secondary potentials in section 5.
\vskip 1cm
\leftline {\bf 5. Secondary potentials}
\vskip 1cm
\noindent
In Ricci-flat 4-manifolds, secondary potentials for spin
${3\over 2}$ are introduced by requiring that locally
[5,15]
$$
\gamma_{A'B'}^{\; \; \; \; \; \; \; C}
\equiv \nabla_{BB'} \; \rho_{A'}^{\; \; \; CB} .
\eqno (5.1)
$$
The insertion of (5.1) into the Rarita-Schwinger equation (A.1)
yields [5,8]
$$
\epsilon_{FL} \; \nabla_{AA'} \;
\nabla^{B'(F} \; \rho_{B'}^{\; \; \; A)L}
+{1\over 2}\nabla_{\; \; A'}^{A} \;
\nabla^{B'M} \; \rho_{B'(AM)}
+\cstok{\ }_{AM} \; \rho_{A'}^{\; \; \; (AM)}
+{3\over 8}\cstok{\ }\rho_{A'}=0
\eqno (5.2)
$$
where $\rho_{A'} \equiv \rho_{A'C}^{\; \; \; \; \; C}$.
Remarkably, equation (5.2) admits a square root in that, if
the following equation holds [5,8,15]:
$$
\nabla^{B'(F} \; \rho_{B'}^{\; \; \; A)L}=0
\eqno (5.3)
$$
then (5.2) reduces to an identity by virtue of spinor Ricci
identities jointly with the basic rules of two-spinor
calculus [8]. However, if the trace-free part of the Ricci
spinor vanishes but $\Lambda$ does not vanish, the effect of
$\Lambda$ makes it necessary to write both (5.3) and the
equation [5]
$$
\rho_{A'}=2{\widetilde \alpha}_{A'}
\eqno (5.4)
$$
where ${\widetilde \alpha}_{A'}$ is a spinor field solving
the Weyl equation [8,16]. An analogous local construction
holds for the $\Gamma$-potentials. The corresponding secondary
potentials are defined locally as
$$
\Gamma_{AB}^{\; \; \; \; \; C'} \equiv
\nabla_{BB'} \; \theta_{A}^{\; \; C'B'} .
\eqno (5.5)
$$
The insertion of (5.5) into the Rarita-Schwinger equation (A.3)
yields a second-order equation whose validity is ensured by the
first-order equation [5]
$$
\nabla^{B(F'} \; \theta_{B}^{\; \; A')L'}=0
\eqno (5.6)
$$
jointly with [5]
$$
\theta_{A}=2\alpha_{A}
\eqno (5.7)
$$
where $\theta_{A} \equiv \theta_{AC'}^{\; \; \; \; \; C'}$,
and $\alpha_{A}$ solves the Weyl equation
$\nabla^{AA'} \; \alpha_{A}=0$ [8,16].

According to the prescription of section 2, which amounts
to replacing $\nabla_{AA'}$ by $S_{AA'}$ in the field
equations [9,10], we now {\it assume} that the super
Rarita-Schwinger equations corresponding to (5.3) and (5.6)
are (see section 7)
$$
S^{B'(F} \; \rho_{B'}^{\; \; \; A)L}=0
\eqno (5.8)
$$
$$
S^{B(F'} \; \theta_{B}^{A')L'}=0
\eqno (5.9)
$$
where the secondary potentials are subject locally to the
supergauge transformations
$$
{\widehat \rho}_{B'}^{\; \; \; AL} \equiv
\rho_{B'}^{\; \; \; AL}+S_{B'}^{\; \; \; A} \; \mu^{L}
\eqno (5.10)
$$
$$
{\widehat \theta}_{B}^{\; \; A'L'} \equiv
\theta_{B}^{\; \; A'L'}+S_{B}^{\; \; A'} \; \zeta^{L'} .
\eqno (5.11)
$$
The analysis of the gauge freedom of the second kind is
entirely analogous to the one in section 3, since equations
like (2.4)-(2.5) now apply to $\mu_{L}$ and $\zeta_{L'}$.
Hence we do not repeat this investigation.

A more general gauge freedom of the twistor type relies on
the supertwistor equations (see (4.12))
$$
S_{B'}^{\; \; \; (A} \; \mu^{L)}
=\nabla_{B'}^{\; \; \; (A} \; \mu^{L)}=0
\eqno (5.12)
$$
$$
S_{B}^{\; \; (A'} \; \zeta^{L')}=\nabla_{B}^{\; \; (A'} \;
\zeta^{L')}=0 .
\eqno (5.13)
$$
Thus, on requiring the preservation of the super Rarita-Schwinger
equations (5.8)-(5.9) under the supergauge transformations
(5.10)-(5.11), one finds the preservation conditions
$$
S^{B'(F} \; S_{B'}^{\; \; \; A)} \; \mu^{L}=0
\eqno (5.14)
$$
$$
S^{B(F'} \; S_{B}^{\; \; A')} \; \zeta^{L'}=0
\eqno (5.15)
$$
which lead to
$$
(f_{1}+f_{2})\pi_{F}+(f_{1}f_{2}-\Lambda)\mu_{F}=0
\eqno (5.16)
$$
$$
(f_{1}+f_{2}){\widetilde \pi}_{F'}
+(f_{1}f_{2}-\Lambda)\zeta_{F'}=0 .
\eqno (5.17)
$$
Hence we can repeat the remarks following equations
(4.16)-(4.19). Again, it is essential that $\pi_{F},\mu_{F}$
and ${\widetilde \pi}_{F'},\zeta_{F'}$ may be unrelated if
(4.18)-(4.19) hold. In the massless case, this is impossible,
and hence there is no gauge freedom compatible with a
non-vanishing cosmological constant [8].

If one does not assume the validity of equations (5.12)-(5.13),
the general preservation equations (5.14)-(5.15) lead instead
to the compatibility conditions
$$
\psi_{\; \; \; \; \; \; D}^{AFL} \; \mu^{D}
-2\Lambda \; \mu^{(A} \; \epsilon^{F)L}
+2f_{2} \omega^{(AF)L}
+f_{1} \epsilon^{L(A} \; T^{F)}
+f_{1} \epsilon^{L(A} \; S^{F)B'} \; \zeta_{B'}=0
\eqno (5.18)
$$
$$
{\widetilde \psi}_{\; \; \; \; \; \; \; \; \; D'}^{A'F'L'}
\; \zeta^{D'} -2\Lambda \; \zeta^{(A'} \; \epsilon^{F')L'}
+2f_{1}{\widetilde \omega}^{(A'F')L'}
+f_{2} \epsilon^{L'(A'} \; {\widetilde T}^{F')}
+f_{2} \epsilon^{L'(A'} \; S^{F')B} \; \mu_{B}=0 .
\eqno (5.19)
$$
If we now combine the compatibility equations (4.20)-(4.23)
with (5.18)-(5.19), and require that the gauge fields
$\nu_{A},\lambda_{A'},\mu_{A},\zeta_{A'}$ should not depend
explicitly on the curvature of the background, we find
that the trace-free part of the Ricci spinor has to
vanish, and the Riemannian 4-geometry is forced to be
conformally flat, since under our assumptions the equations
$$
\psi_{AFLD} \; \mu^{D}=0
\eqno (5.20)
$$
$$
{\widetilde \psi}_{A'F'L'D'} \; \zeta^{D'}=0
\eqno (5.21)
$$
force the anti-self-dual and self-dual Weyl spinors to vanish.
Remarkably, equations (5.20)-(5.21) are just the integrability
conditions for the existence of non-trivial solutions of the
supertwistor equations (5.12)-(5.13). Hence the spinor fields
$\omega,T,{\widetilde \omega}$ and $\widetilde T$ in
(5.18)-(5.19) are such that these equations reduce to
(5.16)-(5.17). In other words, for massive spin-${3\over 2}$
potentials, the gauge freedom is indeed
generated by solutions of the twistor equations in conformally
flat Einstein 4-manifolds.

Last, on inserting the local equations (5.1) and (5.5) into the
second half of the Rarita-Schwinger equations
(cf (A.2) and (A.4) of the appendix), and then replacing
$\nabla_{AA'}$ by $S_{AA'}$ [9], one finds equations whose
preservation under the supergauge transformations (5.10)-(5.11)
is again guaranteed if the supertwistor equations
(5.12)-(5.13) hold.
\vskip 100cm
\leftline {\bf 6. Non-linear superconnection}
\vskip 1cm
\noindent
As a first step in the proof
that (5.8)-(5.9) arise naturally as
integrability conditions of a suitable connection,
we introduce a partial superconnection
$W_{A'}$ (cf [15]) acting on unprimed spinor
fields $\eta_{D}$ defined on the
Riemannian background.

With our notation [15]
$$
W_{A'} \; \eta_{D} \equiv \eta^{A} \; S_{AA'}
\; \eta_{D} - \eta_{B} \; \eta_{C} \;
\rho_{A'}^{\; \; \; BC}
\; \eta_{D} .
\eqno (6.1)
$$
Writing
$$
W_{A'}=\eta^{A} \; \Omega_{AA'}
\eqno (6.2)
$$
where the operator $\Omega_{AA'}$ acts on spinor fields
$\eta_{D}$, we obtain
$$
\eta^{A} \; \Omega_{AA'}
=\eta^{A} \; S_{AA'}-\eta_{B} \; \eta_{C} \;
\rho_{A'}^{\; \; \; BC} .
\eqno (6.3)
$$
Following [15], we require that $\Omega_{AA'}$
should provide a genuine
superconnection on the spin-bundle,
so that it acts in any direction.
Thus, from (6.3) we can take (cf [15])
$$
\Omega_{AA'} \equiv S_{AA'}-\eta^{C} \; \rho_{A'AC}=
S_{AA'}-\eta^{C} \; \rho_{A'(AC)}
+{1\over 2}\eta_{A} \; \rho_{A'} .
\eqno (6.4)
$$
Note that (6.4) makes it necessary to know the trace $\rho_{A'}$,
whilst in (6.1) only the symmetric part of
$\rho_{A'}^{\; \; \; BC}$  survives.
Thus we can see that, independently of the analysis in the
previous sections, the definition of $\Omega_{AA'}$ picks out
a potential of the Rarita-Schwinger type [15].
\vskip 10cm
\leftline {\bf 7. Integrability condition}
\vskip 1cm
\noindent
In section 6 we have introduced a
superconnection $\Omega_{AA'}$ which acts on a bundle
with non-linear fibres, where the term $-\eta^{C} \;
\rho_{A'AC}$ is responsible for the non-linear nature
of $\Omega_{AA'}$ (see (6.4)). Following [15],
we now pass to a description in terms of a vector bundle
of rank three. On introducing the local coordinates
$(u_{A},\xi)$, where
$$
u_{A}=\xi \; \eta_{A}
\eqno (7.1)
$$
the action of the new operator
${\widetilde \Omega}_{AA'}$ reads (cf [15])
$$
{\widetilde\Omega}_{AA'}(u_{B},\xi)
\equiv  \Bigr(S_{AA'} \; u_{B},
S_{AA'} \; \xi-u^{C} \; \rho_{A'AC}\Bigr) .
\eqno (7.2)
$$
Now we are able to prove that (5.8)-(5.9) are integrability
conditions.

The super $\beta$-surfaces are totally null two-surfaces
whose tangent vector has the form $u^{A} \; \pi^{A'}$,
where $\pi^{A'}$ is varying and $u^{A}$ obeys the equation
$$
u^{A} \; S_{AA'} \; u_{B}=0
\eqno (7.3)
$$
which means that $u^{A}$ is supercovariantly constant
over the surface. On defining
$$
\tau_{A'} \equiv u_{B} \; u_{C} \; \rho_{A'}^{\; \; \; BC}
\eqno (7.4)
$$
the condition for ${\widetilde \Omega}_{AA'}$ to be
integrable on super $\beta$-surfaces is (cf [15])
$$
u^{A} \; {\widetilde \Omega}_{AA'} \; \tau^{A'}=
u_{A} \; u_{B} \; u_{C} \;
S^{A'(A} \; \rho_{A'}^{\; \; \; B)C}= 0
\eqno (7.5)
$$
by virtue of the Leibniz rule and of (7.2)-(7.4).
Equation (7.5) implies
$$
S^{A'(A} \; \rho_{A'}^{\; \; \; B)C}=0
\eqno (7.6)
$$
which is the equation (5.8). Similarly, on studying
super $\alpha$-surfaces defined by the equation
$$
{\widetilde u}^{A'} \; S_{AA'} \; {\widetilde u}_{B'}=0
\eqno (7.7)
$$
one obtains (5.9). Thus, although (5.8)-(5.9) are
naturally suggested by the local theory of
spin-${3\over 2}$ potentials, they have a deeper geometric
origin, as shown.
\vskip 1cm
\leftline {\bf 8. Gauge invariance of boundary conditions}
\vskip 1cm
\noindent
In the presence of boundaries one has to impose a suitable
set of boundary
conditions. We study the gauge invariance of
locally supersymmetric boundary conditions first
proposed in [18], which make it possible
to relate bosonic and fermionic fields trough the
action of complementary projection operators at the boundary [8,19].
On using two-component spinor notation for supergravity [20-21],
the spin-$3\over 2$
boundary conditions relevant for quantum cosmology
and supergravity theories are [7,8,21]
$$
\sqrt{2} \; {_{e}n_{A}^{\; \; A'}} \;
\psi_{\; \; i}^{A}= \pm
{\widetilde \psi}_{\; \; i}^{A'}
\; \; \; \; {\rm at} \; \; \; \; \partial M
\eqno (8.1)
$$
where ${_{e}n_{A}^{\; \; A'}}$ is the Euclidean normal to
the boundary [5-8,21] and
$\Bigr(\psi_{\; \; i}^{A},{\widetilde \psi}_{\; \; i}^{A'}\Bigr)$
are the {\it independent} (i.e. not related by any conjugation)
spatial components (hence $i=1,2,3$) of the spinor-valued
one-forms appearing in the action
functional of Euclidean supergravity [20,21].
In terms of the spatial components
$e_{AB'i}$ of the tetrad, and of the primary potentials,
$\Bigr(\psi_{\; \; i}^{A},
{\widetilde \psi}_{\; \; i}^{A'}\Bigr)$ can be
expressed as [3,8,20]
$$
\psi_{A \; i}= \Gamma_{\; \; AB}^{C'}
\; e_{\; \; C'i}^{B} \;
\eqno (8.2)
$$
$$
{\widetilde \psi}_{A' \; i}=
\gamma_{\; \; A'B'}^{C} \;
e_{C \; \; \; i}^{\; \; B'} \; .
\eqno (8.3)
$$
Bearing in mind that the gauge freedom is generated
by solutions of the supertwistor equations (cf (4.1)-(4.2)),
the boundary conditions
(8.1) are preserved under the action
of the supergauge transformations (2.2)-(2.3)
if the spinor fields $\nu^{C}$, $\lambda^{C'}$, $\pi^{C}$
and ${\widetilde \pi}^{C'}$ obey the boundary conditions
$$
\sqrt{2} \; {{}_{e}n_{A}^{\; \; A'}} \;
\Bigr({\widetilde \pi}^{C'} +
f_{1} \; \lambda^{C'}\Bigr)
\; e^{A}_{\; \; C'i}
= \pm \Bigr(\pi^{C} + f_{2}\nu^{C} \Bigr)
e_{C \; \; \; i}^{\; \; A'}
\; \; \; \;  {\rm at} \; \; \; \;
{\partial M} .
\eqno (8.4)
$$
Thus, we have obtained a simple algebraic
relation among the spinor fields
occurring in (4.1)-(4.2), which ensures the gauge invariance of
the boundary conditions (8.1).
\vskip 1cm
\leftline {\bf 9. Concluding remarks}
\vskip 1cm
\noindent
We have given an entirely two-spinor description of massive
spin-${3\over 2}$ potentials in Einstein 4-geometries.
Although the supercovariant derivative
(2.1) was well-known in the literature, following the work
in [9], and its Lorentzian version was already applied in
[13,17], the systematic analysis of primary and secondary
potentials with the local form of their supergauge
transformations was not yet available in the literature, to
the best of our knowledge.

Our first result is the two-spinor
proof that, for massive spin-${3\over 2}$ potentials, the
gauge freedom is generated by solutions of
the supertwistor equations in conformally flat Einstein
4-manifolds. Moreover, we have shown that the first-order
equations (5.8)-(5.9),
whose consideration is suggested by the local
theory of massive spin-${3\over 2}$ potentials, admit a
deeper geometric interpretation as integrability
conditions on super $\beta$- and super $\alpha$-surfaces
of a connection on a rank-three vector bundle.
This result generalizes the analysis of massless
spin-${3\over 2}$ fields appearing in [15].
Besides that, in the presence of boundaries we
have found the condition under
which locally supersymmetric boundary conditions [18-19]
are gauge-invariant. One now has to find explicit solutions
of the equations (2.9)-(2.12), and the supercovariant form
of $\beta$-surfaces studied in our paper deserves
a more careful consideration.
Hence we hope that our work can lead to a better understanding
of twistor geometry and consistent supergravity theories
in four-dimensions.
\vskip 1cm
\leftline {\bf Appendix}
\vskip 1cm
\noindent
For completeness, we write the Rarita-Schwinger equations
for massless spin-${3\over 2}$ potentials in Ricci-flat
4-manifolds. They take the form [1-5]
$$
\epsilon^{B'C'} \; \nabla_{A(A'} \;
\gamma_{\; \; B')C'}^{A}=0
\eqno (A.1)
$$
$$
\nabla^{B'(B} \; \gamma_{\; \; \; B'C'}^{A)}=0
\eqno (A.2)
$$
$$
\epsilon^{BC} \; \nabla_{A'(A} \;
\Gamma_{\; \; \; B)C}^{A'}=0
\eqno (A.3)
$$
$$
\nabla^{B(B'} \; \Gamma_{\; \; \; \; BC}^{A')}=0 .
\eqno (A.4)
$$
Note that, if one works with $\nabla_{AA'}$ when $\Lambda$
does not vanish, the right-hand sides of (A.1) and (A.3)
should be replaced by $-3\Lambda \; {\widetilde \alpha}_{A'}$
and $-3\Lambda \; \alpha_{A}$ respectively, where
$\alpha_{A}$ and ${\widetilde \alpha}_{A'}$ are spinor fields
solving the Weyl equations [5,8,16]. In the massless case
$\Lambda$ is forced to vanish [1-4,8], but for massive models such
contributions should be taken into account (cf our equations
(2.9) and (2.11)).

In [13], the equation for Lorentzian Killing spinors is
written in the form (see also equations (29) of [17], and cf
our equations (3.1)-(3.2))
$$
\nabla_{AX'}O_{B}=b \; \epsilon_{AB} \; {\overline O}_{X'}
\eqno (A.5)
$$
where the parameter $b$ is proportional to $\sqrt{-\Lambda}$,
and the overbar denotes, as usual, the complex conjugation
of spinors.
\vskip 1cm
\leftline {\bf Acknowledgments}
\vskip 1cm
\noindent
It is a pleasure to thank Paul Townsend for a useful
conversation on massive spin-${3\over 2}$ potentials. Our
research was made possible in part by the European Union
under the Human Capital and Mobility programme.
\vskip 10cm
\leftline {\bf References}
\vskip 1cm
\item {[1]}
Penrose R 1991 Twistors as Spin-${3\over 2}$ Charges
{\it Gravitation and Modern Cosmology} eds A Zichichi, V
de Sabbata and N S\'anchez (New York: Plenum Press)
\item {[2]}
Penrose R 1991 {\it Twistor Newsletter} {\bf 32} 1
\item {[3]}
Penrose R 1991 {\it Twistor Newsletter} {\bf 33} 1
\item {[4]}
Penrose R and Mason L J {\it Twistor Newsletter}
{\bf 37} 1
\item {[5]}
Esposito G 1995 {\it Complex General Relativity}
(Fundamental Theories of Physics {\bf 69})
(Dordrecht: Kluwer)
\item {[6]}
Esposito G and Pollifrone G 1994 {\it Class. Quantum Grav.}
{\bf 11} 897
\item {[7]}
Esposito G and Pollifrone G 1994 Twistors and Spin-${3\over 2}$
Potentials in Quantum Gravity {\it Twistor Theory} ed S Huggett
(New York: Marcel Dekker)
\item {[8]}
Esposito G, Gionti G, Kamenshchik A Yu, Mishakov I V and
Pollifrone G {\it Spin-${3\over 2}$ Potentials in Backgrounds
with Boundary} (DSF preprint 95/20)
\item {[9]}
Townsend P K 1977 {\it Phys. Rev.} D {\bf 15} 2802
\item {[10]}
Izquierdo J M and Townsend P K 1995 {\it Class. Quantum Grav.}
{\bf 12} 895
\item {[11]}
Penrose R and Rindler W 1986 {\it Spinors and Space-Time, Vol. 2:
Spinor and Twistor Methods in Space-Time Geometry}
(Cambridge: Cambridge University Press)
\item {[12]}
Penrose R and Rindler W 1984 {\it Spinors and Space-Time, Vol. 1:
Two-Spinor Calculus and Relativistic Fields} (Cambridge:
Cambridge University Press)
\item {[13]}
Siklos S T C 1985 Lobatchevski Plane Gravitational Waves
{\it Galaxies, Axisymmetric Systems and Relativity} ed
M A H MacCallum (Cambridge: Cambridge University Press)
\item {[14]}
Lewandowski J 1991 {\it Class. Quantum Grav.} {\bf 8} L11
\item {[15]}
Penrose R 1994 Twistors and the Einstein Equations
{\it Twistor Theory} ed S Huggett (New York: Marcel Dekker)
\item {[16]}
Aichelburg P C and Urbantke H K 1981 {\it Gen. Rel. Grav.}
{\bf 13} 817
\item {[17]}
Perry M J 1984 The Positive Mass Theorem and Black Holes
{\it Asymptotic Behaviour of Mass and Spacetime Geometry}
ed J Flaherty (Berlin: Springer)
\item {[18]}
Luckock H C and Moss I G 1989 {\it Class. Quantum Grav.}
{\bf 6} 1993
\item {[19]}
Luckock H C 1991 {\it J. Math. Phys.} {\bf 32} 1755
\item {[20]}
D'Eath P D 1984 {\it Phys. Rev} D {\bf 29} 2199
\item {[21]}
Esposito G 1994 {\it Quantum Gravity, Quantum Cosmology and
Lorentzian Geometries} (Lecture Notes in Physics {\bf m12})
(Berlin: Springer)

\bye